\documentclass[final,5p]{elsarticle}

\usepackage{amssymb}

\usepackage[utf8]{inputenc}
\usepackage{graphicx}
\usepackage{rotating}
\usepackage{mathptmx}
\usepackage{comment} 
\usepackage{xcolor}
\usepackage{hyperref}

\journal{Nucl. Instrum. Methods Phys. Res. A: Accel. Spectrom. Detect. Assoc. Equip.}

\begin{document}

\begin{frontmatter}

\date{May 2022}



\title{Radiation-induced secondary emissions in solid-state devices \\ as a possible contribution to quasiparticle poisoning of superconducting circuits}


\author[PNNL]{Francisco Ponce\corref{cor1}}
\ead{francisco.ponce@pnnl.gov}
\cortext[cor1]{Corresponding author}

\author[PNNL]{John L. Orrell}
\author[PNNL]{Zheming Wang}

\affiliation[PNNL]{organization={Pacific Northwest National Laboratory},
            addressline={902 Battelle Blvd.}, 
            city={Richland},
            postcode={99352}, 
            state={WA},
            country={U.S.A.}}

\begin{abstract}
This report estimates the potential for secondary emission processes induced by ionizing radiation to result in the generation of quasiparticles in superconducting circuits. These estimates are based on evaluation of data collected from a small superconducting detector and a fluorescence measurement of typical read-out circuit board materials. Specifically, we study cosmic ray muons interacting with substrate or mechanical support materials present within the vicinity of superconducting circuits. We evaluate the potential for secondary emission, such as scintillation and/or fluorescence, from these nearby materials to occur at sufficient energy (wavelength) and rate (photon flux) to ultimately lead to the breaking of superconducting Copper pairs (\textit{i.e.}, production of quasiparticles). This evaluation leads to a conclusion that material fluorescence in the vicinity of superconducting circuits is a potential contributor to undesirable elevated quasiparticle populations. A co-design approach evaluating superconducting circuit design and the material environment within the immediate vicinity of the circuit would prove beneficial for mitigating undesired environmentally-induced influences on superconducting device performance, such as in direct detection dark matter sensors or quantum computing bits (qubits).
\end{abstract}



\end{frontmatter}



%
%

\section{Introduction}
Ionizing radiation backgrounds have been observed to effect superconducting quantum bit (qubit) coherence times~\cite{Vepsalainen2020} as well as superconducting resonators~\cite{Cardani2021}. These effects involve highly energetic particles depositing energy in the superconducting circuit and/or its substrate, leading to quasiparticle poisoning, which can cause correlated errors across an entire qubit array~\cite{Wilen2021, McEwen2022}. Another related possibility is that qubits are affected by secondary emissions~\cite{Du2022}. Specifically, secondary emissions --- transition radiation, Cherenkov radiation, and/or fluorescence --- may contribute to quasiparticle production in superconducting circuits. As highly energetic particles pass through the dilution refrigerator (DR) and qubit device packaging, these secondary emissions may be generated near the device which may result in quasiparticle production and lead to qubit decoherence.

Currently, secondary emission processes are under consideration by the direct detection dark matter (DM) research community as a potential explanation of low energy ($\sim$keV scale) event excesses observed in multiple DM experiments~\cite{Excess2022}. The printed circuit board (PCB) and other insulating materials used for device readout are potential culprits. In this respect, qubit devices and DM detectors share a commonality; PCBs are typically used to interface readout wiring with the circuitry patterned onto the device substrates. The PCB materials are believed to fluoresce in the optical and ultraviolet (UV) ranges of the light spectrum with lifetimes ranging from ns to 10’s of ms~\cite{Trukhin2003}. Corroborating evidence for this was recently shown in the SuperCDMS HVeV detector operated near sea level at Stanford University~\cite{Ponce2022}. The experiment made use of optical (650 nm) and UV (275 nm) photons to demonstrate a controllable fluorescence spectrum inside the DR. The response to these controlled optical and UV emissions is consistent with the secondary emission hypothesis. The photon energies are sufficient to partially explain the low-energy event excesses observed in various DM detectors (see, e.g., Ref.~\cite{SCDMS2022}). If these secondary emission processes are implicated in DM detectors, then they may also contribute to quasiparticle production in superconducting qubit circuits. 

Assuming the above hypothesis is correct, in this paper we outline a simple model and estimate the efficiency for secondary-emission photons to reach a device (assuming no reflections), using the SuperCDMS HVeV device setup in Ref. ~\cite{romani,hvevRun1,Ponce2022} as a concrete example of physical device layout. We use the properties of NaI and CdWO$_4$ scintillator crystals---\emph{as an extreme case}---in place of the PCB where properties are not available to set an upper bound on the expected secondary emission. We further compare our results to those of Ref.~\cite{Du2022}. Furthermore, we measure the fluorescence of the G10 PCB used in the Ref.~\cite{romani,hvevRun1,Ponce2022} experiments and confirm that the output is consistent with fluorescence measurements in the literature ~\cite{Spizzichino2016}. Finally, we compare the fluorescence properties of other materials used to interface with dark matter detectors and qubit devices.
\begin{figure}
    \centering
    \begin{minipage}{0.49\textwidth}
        \begin{center}
        \includegraphics[width=1\textwidth]{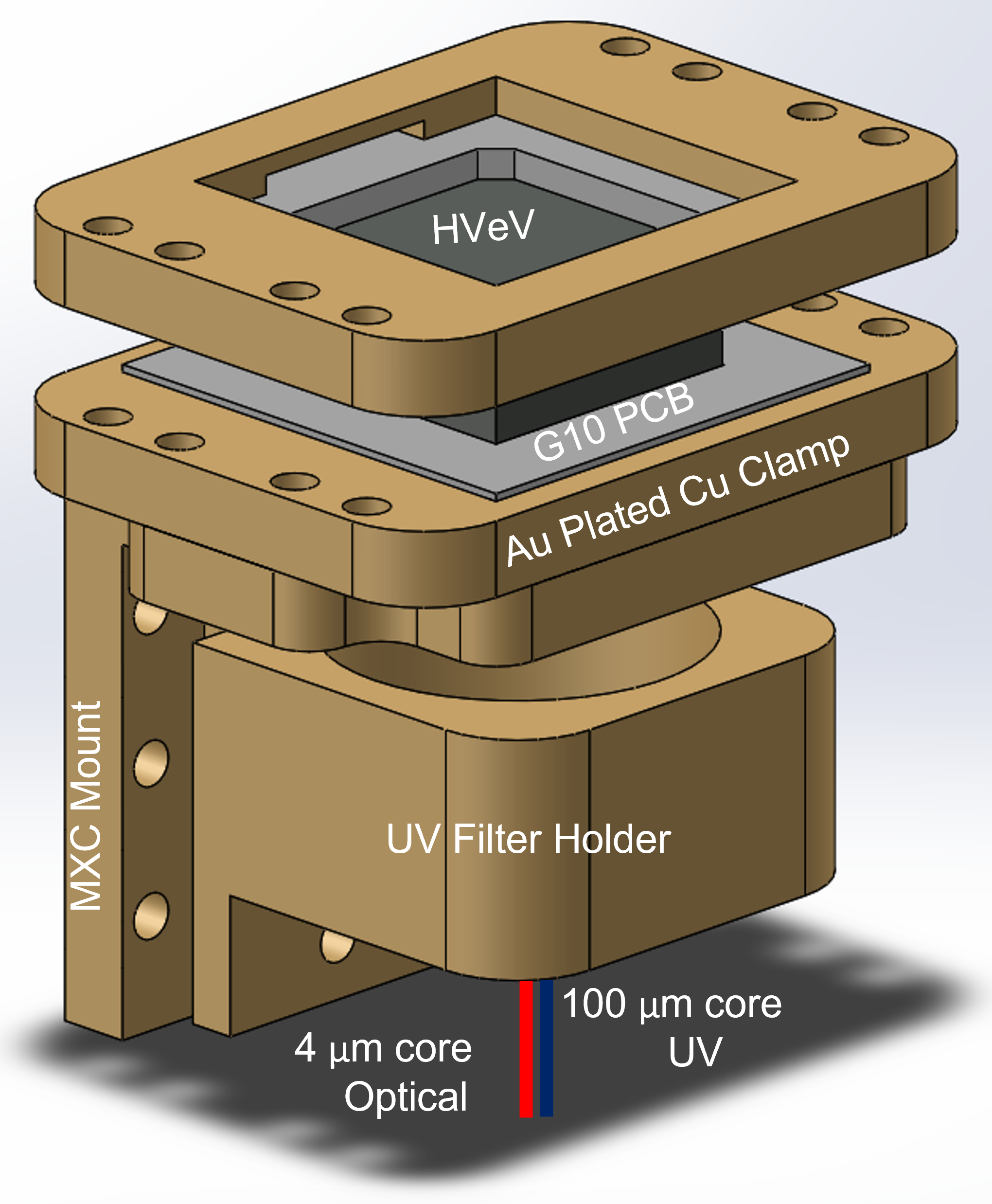}
        \caption{CAD rendering of the HVeV device setup used to study UV photons in Ref.~\cite{Ponce2022}.}
        \label{SU_Setup}
        \end{center}
    \end{minipage}\hfill
    \begin{minipage}{0.49\textwidth}
        \begin{center}
        \includegraphics[width=1\textwidth]{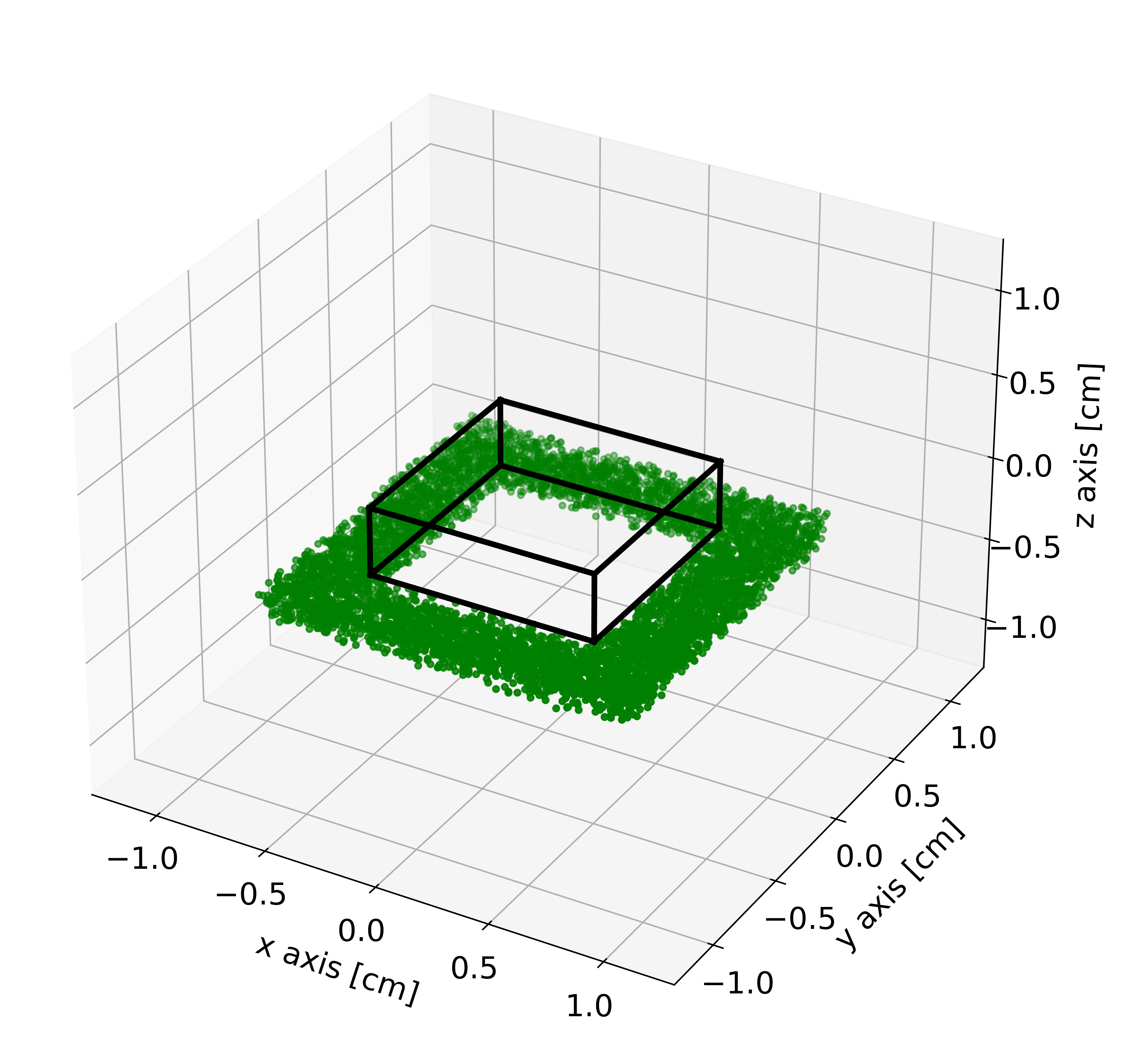}
        \caption{Monte Carlo simulation of the generation of photons throughout the lower G10 PCB relative to the detector volume (black box) with the geometry cut applied to exclude simulated events in the machined square hole area.}
        \label{MC}
        \end{center}
    \end{minipage}
\end{figure}

\section{G10 PCB Fluorescence Incidence Model}
Recent work made use of optical (650~nm) and UV (275~nm) photons to demonstrate a controllable fluorescence spectrum inside a DR~\cite{Ponce2022}. The spectrum was measured by the SuperCDMS HVeV DM detector shown in Fig.~\ref{SU_Setup}, which consists of a superconducting circuit patterned onto a silicon substrate. The optical photons were generated from a pulsed laser coupled to a 4~$\mu$m core single-mode fiber optic (FO), while the UV photons were generated from a pulsed high power UV LED coupled to a 100~$\mu$m core multi-mode FO designed to transmit UV radiation without fluorescing. The laser and LED were synchronized using a delay generator so that the trigger from the LED would initialize the laser. This setup enabled the HVeV detector to be illuminated in alternation by optical and UV photons, each at a rate of 50~Hz for a combined rate of 100~Hz with 10~ms time spacing between alternating optical and UV photon illuminations. In Ref.~\cite{Ponce2022}, the fluorescence was demonstrated to occur only after pulsing the UV LED. As the fluorescence was controllable via UV LED excitation, this is suggestive that external high energy injections may also excite these fluorescing states. What is specifically unknown is the micro-physical process for production of fluorescence photons for PCB materials, such as G10.

To perform a quantitative evaluation, we construct a geometric model of the HVeV detector setup that was operated at Stanford University at an elevation of ~25 m above sea level. The device was operated with minimal overburden (only the building) and no intentional shielding within the laboratory. The HVeV detector was sandwiched between two G10 PCB boards using Cu clamps to hold everything together, as illustrated in Fig.~\ref{SU_Setup}. The detector was orientated with its athermal phonon sensors---based on transition edge sensors (TESs)~\cite{Irwin1995}---facing up ($+z$ axis) and the Al parquet pattern (used to create a controlled voltage plane) facing down ($-z$ axis). A 275~nm UV band-pass filter was positioned in front of the LED at room temperature and a second one was held in a Cu holder via a plastic ring inside the DR at $\sim$30~mK. Attached to the Cu holder, below the filter and inside the DR, are two FOs separated by 1~mm used in alternation to illuminate the parquet side of the device with optical and/or UV photons through the UV filter, as described above. The Cu clamp, Au-plated Cu parts, and brass bolts were excluded as potential fluorescence sources, because these materials have significantly attenuated fluorescence yields~\cite{Hubbell1994}. The G10 PCB and the UV filter's plastic ring are the key suspects for any secondary fluorescence emission.

The G10 PCB is a 0.152~cm  thick square of side length 1.651~cm. Two centered square holes are machined into the G10 PCB. The smaller is a through-hole of side length 0.980~cm. The larger hole has a depth of 0.051~cm with a side length of 1.016~cm to center. The larger hole physically centers the HVeV detector side-to-side. The G10 PCB is used to clamp the detector from above and below as depicted in Fig. \ref{SU_Setup}. 

There are several potential sources of primary ionizing radiation that could excite fluorescence, including X-rays, $\gamma$-rays, and muons. The X-rays and $\gamma$-rays in the environment of the Stanford laboratory have not been well-characterized. On the other hand, muons have a well-defined energy deposition profile dependent on the material density, and they have a nominally constant flux for a fixed altitude. Furthermore, the muon rate can be compared from location to location, taking altitude into account, potentially making direct comparisons between laboratories feasible. Given these existing conditions, we focus on the muon flux at (or near) sea level as our primary excitation of fluorescence states in the model.

The muons ($\mu$) at sea level have a mean energy of 4~GeV and a flux of 0.019~$\mu$/cm$^2$/s~\cite{Jovanovic2018}. The generalized average energy loss of a muon as it traverses a material is $dE/dx = 2$~MeV/(cm$\cdot$g/cm$^3$), which can be scaled to different materials based on density. This is effectively a minimum ionizing particle interaction model. For the purposes of this paper, we focus solely on energy deposited from muons in the G10 PCBs holding the HVeV device. 

The density of G10 PCB is taken to be $\rho_{G10}$ = 1.7~g/cm$^3$~\cite{pcbDensity} (\textit{i.e.}, the density for FR4) which sets the expected average energy deposited by a muon traversing the board to be 0.518~MeV, assuming a minimum ionizing particle model. We assume that the total deposited energy goes into generating photons across all wavelengths proportionally to the G10 PCB fluorescence spectrum. We limit our analysis to fluorescence wavelengths below 1127~nm ($\geq$~1.1 eV)~\cite{Kittel1986} as these photons have sufficient energy to break at least one electron-hole pair in silicon. For simplicity, we consider photons that break no more than one e-h pair (on average), which sets a lower wavelength bound of 326~nm ($\leq$~3.8~eV)~\cite{Hodgkinson1963}. Once these constraints and efficiencies are accounted for, the absorbed portion of the fluorescence spectrum is integrated over the range of relevant wavelengths (326~nm $\leq \lambda \leq$ 1127~nm) to calculate the total e/h pair production rate for a typical HVeV-like silicon chip-based device. 

We calculate the total energy deposited by a muon traveling normal to the PCB face to be $0.518$~MeV/$\mu$. The muon flux through a single G10 PCB is $0.032$~$\mu$/s (with the center square area removed). Thus, we can expect an average contribution of $0.017$~MeV/s of energy from muons. Assuming all of this energy goes into generating photons and that the G10 PCB is no better than the best NaI scintillator crystals~\cite{Jovanovic2018} that generate 45,000~photons/MeV, we can estimate a maximum fluorescence from both G10 PCBs of 1501~photons/s into a 4$\pi$ solid angle. We note that this is a presumptive \emph{maximum} that likely over-estimates the case in fact. In other words, we are evaluating a ``worst case'' scenario, which if nevertheless is shown to be small, could allow us to rule out the fluorescence secondary emission hypothesis. In comparison, CdWO$_4$---one of the lower yielding scintillators in use---would generate 7800~photons/MeV, and the resulting flux from such a material due to muons would be 260~photons/s into a 4$\pi$ solid angle.

To approximate the incident photon flux on the HVeV detector surfaces, we performed a Monte Carlo (MC) simulation of the PCB-detector geometry assuming:  (1) the G10 PCB does not self-absorb any secondary emission photons, (2) the 1$\times$1~cm$^2$ square faces of the detector are excluded, and (3) photons are only absorbed at the first surface they are incident on (\textit{i.e.}, reflections are assumed to be lost). The first assumption is consistent with setting an upper bound on expected rates as it sets the active volume of PCB contributing to secondary emission photons to the maximum possible. The second and third assumptions approximate the fact that the 1$\times$1~cm$^2$ detector faces have direct solid angle to a relatively small portion of PCB; most secondary emission photons would need to successfully reflected at least once to reach a face. The third assumption is also intended to reduce complexity of the model, and is reasonable considering the metal (Cu) package would tend to reflect photons away from the detector.

For the MC we focused on the bottom G10 PCB geometry because the geometry is symmetric about the $xy$ plane and results will be half the total. The python numpy module is used to simulate event locations across the $xy$ plane between $\pm$0.65~cm (in each direction) and between $-0.2$ and $-0.3524$~cm along the $z$ axis. A geometry cut is applied to remove simulated events with positions immediately below the HVeV detector ($\pm$0.5~cm in both directions of the $xy$ plane) to where the G10 PCB is machined (see Fig.~\ref{MC}). The percentage of events passing this cut is 63.3~$\pm$~0.5~\%, which is consistent with the ratio of machined G10 PCB to full area: $100\cdot{}(1.651^2 - 1^2)/1.651^2 = 63.31$\%.

A photon with a random normalized momentum vector was generated at each simulated event location; no energy is assigned to the photon as we are solely interested in the incident flux on the detector. A set of parametric equations were used to determine the path of the simulated photon,
\begin{eqnarray}
x(t)&=& x_i + \dot{x}_it\\
y(t)&=& y_i + \dot{y}_it\\
z(t)&=& z_i + \dot{z}_it
\end{eqnarray}
where $x_i$, $y_i$, and $z_i$ were the Cartesian position of the simulated event and $\dot{x}_i$, $\dot{y}_i$, and $\dot{z}_i$ were the corresponding simulated normalized momentum. We determine which of the detectors four sidewall faces (+0.5~$x$, -0.5~$x$, +0.5~$y$, -0.5~$y$) the photon is incident on (if any) by calculating the time when the parametric equations in the $x$ ($y$) direction equal $\pm$~0.5~cm and then using that time to check if the remaining two dimensions are within $\pm$~0.5~cm in the $y$ ($x$) direction and $\pm$~0.2~cm in the $z$ direction. To prevent double counting, an order cut is applied to remove events that would physically pass through another crystal plane face prior to the plane in question. Finally, simulated events whose time of intercept with the plane in question is negative are discarded as they have momenta in the direction away from the crystal. 

A total of 1000 MC simulations were performed with 10$^4$ events generated in each simulation. On average, the +$\hat{x}$, -$\hat{x}$, +$\hat{y}$, and -$\hat{y}$ planes observed 143.6~$\pm$~11.8, 143.1~$\pm$~12.1, 142.7~$\pm$~11.7, and 143.1~$\pm$~11.9 incident photons out of 6330.4 $\pm$~48.4 simulated photons, respectively. This results in a 8.9 $\pm$~0.4\% probability that a randomly generated photon in the lower G10 PCB volume will be incident on the detector ignoring the effects of the photon attenuation length in the PCB. The attenuation length of the G10 PCB can be readily introduced by an appropriate scaling of the above probability to account for active volume.

Now assuming that the PCB was made from NaI or CdWO$_4$, the PCB would generate $\sim$1500 or $\sim$260~photons/sec, respectively, due to muons traversing the material. Based on this and our MC, we get an incident photon flux of $\sim$267~photons/sec (for NaI) or $\sim$46~photons/sec (for CdWO$_4$) on the detector sidewalls. This is under the strong assumption that the PCB material is transparent to the fluorescence photons. This assumption is appropriate for NaI and CdWO$_4$, but unlikely correct for the actual G10 PCB. We remind the reader again the calculation using NaI and CdWO$_4$ is intended to set an upper bound on possible fluorescence emission.

From the model in Ref.~\cite{Du2022}, we would expect an average of 2.8$\cdot10^4$ Cherenkov events/g$\cdot$day at the surface with this HVeV setup, or 0.302 events/sec. Based on  Eq. C2 in Ref.~\cite{Du2022}, an average of $\sim$223 photons/event are generated (for a muon going through 1.52 mm thick PCB) distributed over a range of frequencies from 1~eV to just above 8~eV. This would result in the two G10 PCBs  fluorescing at a rate of $\sim$135 photons/sec. Given the geometric efficiency from our MC, the HVeV detector would observe an incident flux of $\sim$12~photons/sec from the G10 PCB (treating it as FR-4).

Recall, in our MC simulation we ignored reflectivity, but we note that the reflectivity of Si is between 33\% at 650~nm and 60\% at 300~nm~\cite{Jellison1982}. The fluorescence spectrum of a typical fiberglass material is between 300~nm and 600~nm~\cite{Spizzichino2016}, which is within the single electron-hole pair generation energy for Si. As such the actual incident photon flux absorbed will be $\sim$50\% of the available flux calculated here.

\section{PCB Material Fluorescence}
In considering the above hypothesis and to improve upon existing models, we sought to measure the fluorescence properties of typical materials used in various cryogenic experiments: (1) R04035B~\cite{Huang2021}, (2) TMM10 \cite{Lienhard2019,Dewes2012,Brehm2021}, (3) Cirlex~\cite{PhysRevD.95.082002}, (4) G10 PCB~\cite{hvevRun1}, (5) typical PCB~\cite{hvevRun2}. The materials were mounted in a Horiba Fluorolog III fluorescence spectrometer and exposed to a broad range of excitation wavelengths from a Xe plasma lamp to observe the resulting fluorescence at room temperature. The system is capable of performing both emission scans (300-1300~nm with different gratings) and excitation scans (260-750~nm). The grating is a physical barrier to the Xe lamp, but at extreme settings it allows light to leak through at the edges. This lead to an increased intensity at the edges of our measurements. Thus the spectra presented here were cut off at the edges where the limitations of the grating start to let unfiltered light through. Additionally, we normalize the intensity of the individual emission/excitation scans to the maximum over the entire scan so that they may be plotted together.

In an emission scan, the excitation wavelength $\lambda_{Ex}$ from the light source is held constant while illuminating the sample. The sample fluorescence is then measured as a function of the emitted wavelength, $\lambda_{Em}$. For an excitation scan, the fluorescence is monitored at a fixed $\lambda_{Em}$, while $\lambda_{Ex}$ is varied. For the case of excitation by ionizing radiation, the emission scan is likely the more relevant approach. However, we report all measurement results available to us.

We identify two distinct orientations of the materials: face-normal (area with circuit layout) and edge-normal (area with no circuit layout). Samples were measured at multiple locations on both the face and edge. There was otherwise no special treatment/handling of the materials. The RO4035B and TMM10 were measured only in the edge-normal orientation because the faces were entirely covered by Cu. No differences were observed between face and edge measurements beyond an overall intensity level difference for the other materials. 

\begin{figure}
    \centering
    \begin{minipage}{0.49\textwidth}
        \begin{center}
        \includegraphics[width=1\textwidth]{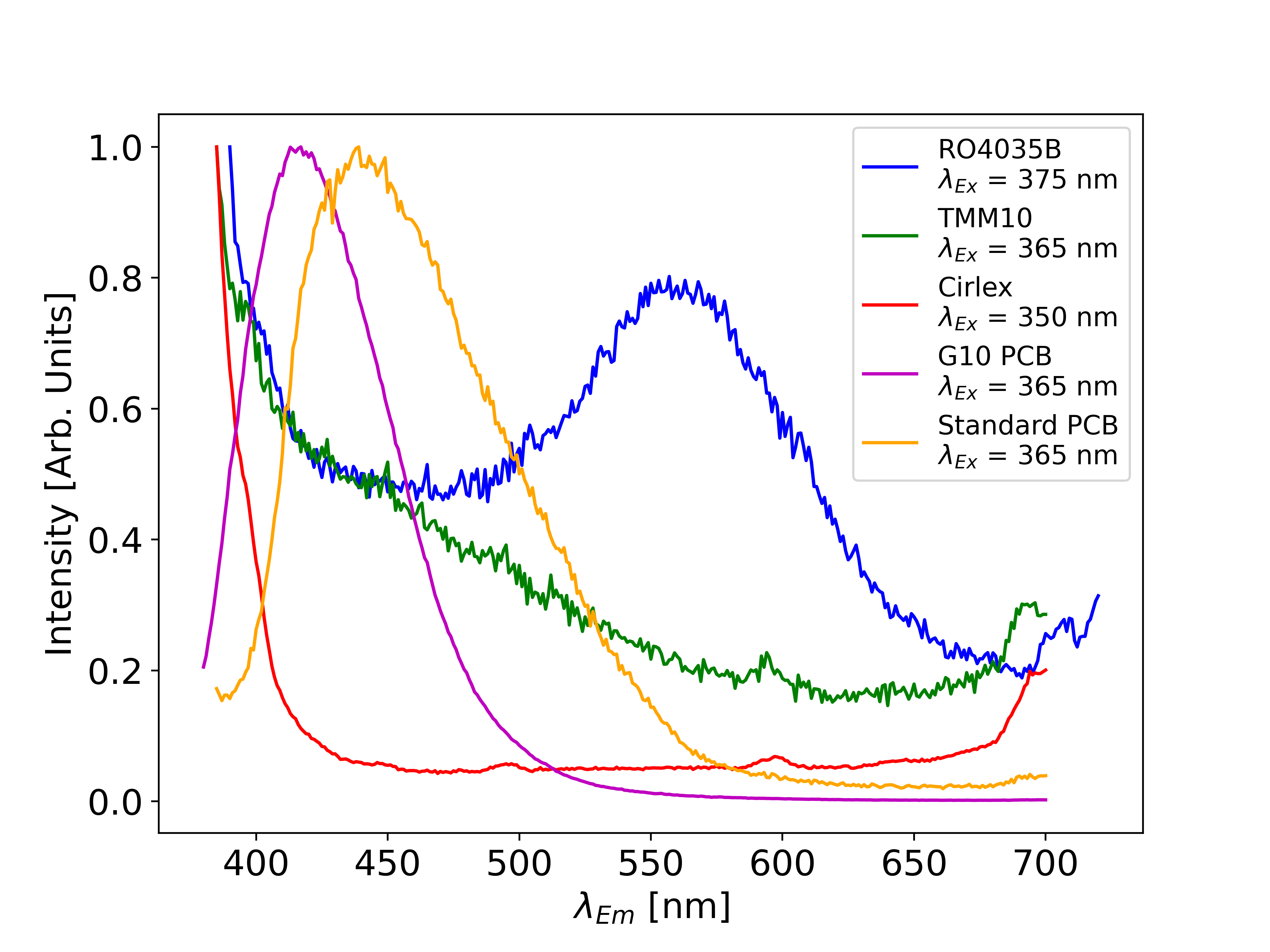}
        \caption{Emission scans were performed at excitation wavelengths which resulted in the strongest observable fluorescence response. Each spectrum is normalized to the maximum for the measurement. The edge effects on the left and right hand sides of the spectra are caused by emission light bypassing the edge of the diffraction grating. The small peaks at 490~nm and 590~nm in the TMM10 and Cirlex are from the Xe lamp.}
        \label{emission}
        \end{center}
    \end{minipage}\hfill
    \begin{minipage}{0.49\textwidth}
        \begin{center}
        \includegraphics[width=1\textwidth]{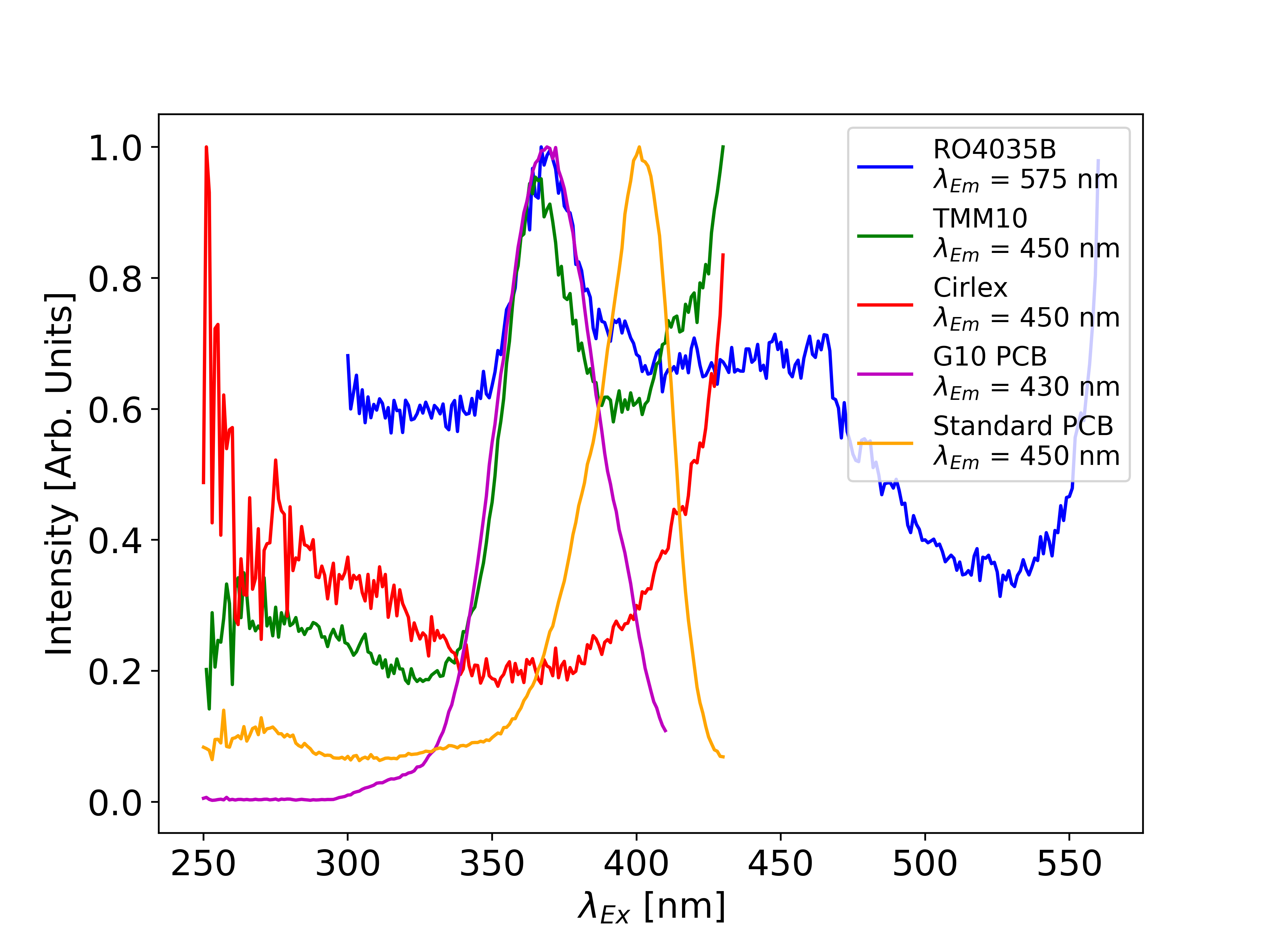}
        \caption{Excitation scans were performed while monitoring the emission wavelength corresponding to the strongest fluorescence response and away from edge effects. Each spectrum is normalized to the maximum for the measurement. The edge effects on the left and right hand sides of the spectra are caused by excitation light bypassing the edge of the diffraction grating.}
        \label{excitation}
        \end{center}
    \end{minipage}
\end{figure}

We measured the RO4035B sample (0.062$^{\prime\prime}$ thick) in the edge-normal orientation. For the emission spectrum, Fig.~\ref{emission} (blue curve), the excitation wavelength was 375~nm. We observed a strong fluorescence at $\sim$560~nm with a broad shoulder at lower wavelengths. For the excitation spectrum, Fig.~\ref{excitation} (blue curve), we monitored the fluorescence at 575~nm. We observed a strong fluorescence just below 375~nm with a broad flat background from 300~nm to $\sim$460 nm at 60-70\% of maximum intensity. Based on these results, RO4035B would convert UV photons from other secondary sources to 560~nm photons and fluoresce at 560~nm when excited by external sources.

We measured the TMM10 sample (0.25$^{\prime\prime}$ thick) in the edge-normal orientation. For the emission spectrum, Fig.~\ref{emission} (green curve),the excitation wavelength was 365~nm. We observed a weak fluorescence across a broad range. The small peak at 590~nm is from the Xe plasma lamp and visible across multiple materials. For the excitation spectrum, Fig.~\ref{emission} (green curve), we monitored the fluorescence at 370~nm. We observed a reflection peak at the excitation wavelength. From this we can conclude that TMM10 would not convert secondary photons into a narrow band and the fluorescence from cosmic rays would likely be broad band.

We measured the Cirlex sample (0.062$^{\prime\prime}$ thick) in both the edge- and face-normal orientations.The Cirlex sample had no additional (non-Cirlex) material on its face. For the emission spectrum, Fig.~\ref{emission} (red curve), the excitation wavelength was 350~nm. We observe two peaks at 490 and 590~nm from the Xe plasma lamp. For the excitation spectrum, Fig.~\ref{excitation} (red curve), we monitored the fluorescence at 450~nm. We see a relatively flat response. Cirlex fluoresces very weakly when illuminated across a broad range of wavelengths and does not appear to reflect UV photons.

We measured the G10 PCB sample (0.06$^{\prime\prime}$ thick) in both the edge- and face-normal orientations.The G10 PCB sample had a sparse Cu pattern on its face. For the emission spectrum, Fig.~\ref{emission} (magenta curve), the excitation wavelength was 365~nm. We observed a strong fluorescence peak at $\sim$425~nm. For the excitation spectra, Fig.~\ref{excitation} (magenta curve), we monitored the fluorescence at 430~nm. We observed that the fluorescence peaked at an excitation wavelength of 370~nm. As such G10 PCB would convert a narrow band of UV photons from other secondary sources to 425~nm photons and fluoresce at 425~nm when excited by cosmic rays.

We measured the standard FR-4 PCB sample (0.062$^{\prime\prime}$ thick) in both the edge- and face-normal orientations. The FR-4 PCB sample had the common glossy green finish of typical PCB boards and sparse exposed Au plated Cu pads on the face. For the emission spectrum, Fig.~\ref{emission} (yellow curve), the excitation wavelength was 365~nm. We observe a strong fluorescence peak at 440~nm with a long high-wavelength tail. For the excitation spectrum, Fig.~\ref{excitation} (yellow curve), we monitored the fluorescence at 450~nm. We observed that the fluorescence peaked at an excitation wavelength of 390~nm with a very broad tail down to 270~nm. As such FR-4 PCB would convert a wide range of UV photons from other secondary sources to 450~nm photons and fluoresce over a broad range of wavelengths from 400~nm to 600~nm when excited by cosmic rays.

\section{Conclusion}
We prepared a simple Monte Carlo model to simulate the geometry of the sandwiched setup used for the measurements in Refs.~\cite{romani,hvevRun1,Ponce2022}. The goal of the model was to help inform a hypothesis regarding whether PCB material fluorescence could generate sufficient photons to lead to e/h pair production (and subsequent quasiparticle production) in superconducting circuits and sensors patterned on Si substrates. Based on this model, fluorescence photons generated in the G10 PCB volume from cosmic rays have an 8.9\% probability of being incident on the HVeV Si-substrate sidewalls. 

We performed fluorescence measurements of five materials used in either dark matter search detectors or quantum information science devices. Among the materials used for dark matter detectors, the Cirlex was observed to not fluoresce or reflect UV photons, whereas the G10 and FR-4 PCBs generally converted UV photons into optical photons. For the quantum information science device materials, TMM10 was not observed to fluoresce, although it would reflect UV from exposed surfaces. The RO4035B was observed to convert a broad band of UV photons into optical photons. 

Under the strong assumption about the G10 PCB properties used in the MC we set an upper bound of $\sim$12 photons/sec based on the model from Ref. ~\cite{Du2022}.Typical materials used in interfacing with the devices down convert UV photons into optical photons. The resulting fluorescence has sufficient energy to break only single $e^{-}h^{+}$-pairs, which is consistent with the high rates of low energy events observed in dark matter experiments and we can expect this to be true for quantum devices. Thus, materials used to interface with devices should be minimized so as to mitigate the potential effects of cosmic rays inducing secondary emission visible to the DM detectors and/or qubits. 

%
%
\section*{Acknowledgements}
We thank Patrick Harrington and the SuperCDMS collaboration for providing the materials studied in this research. This material is based upon work supported by the U.S. Department of Energy, Office of Science, National Quantum Information Science Research Centers, Co-design Center for Quantum Advantage (C2QA) under contract number DE-SC0012704. PNNL is operated by the Battelle Memorial Institute for the U.S. Department of Energy under contract DE-AC05-76RL01830.

\bibliographystyle{elsarticle-num}
\bibliography{References}

\end{document}